\documentclass[conference]{IEEEtran}

\IEEEoverridecommandlockouts
\usepackage{cite}
\usepackage{amsmath,amssymb,amsfonts}
\usepackage{algorithmic}
\usepackage{graphicx}
\usepackage{textcomp}
\usepackage{xcolor}
\usepackage[left=1.6002cm,right=1.6002cm,top=1.905cm,bottom=2.54cm]{geometry}

\def\BibTeX{{\rm B\kern-.05em{\sc i\kern-.025em b}\kern-.08em
    T\kern-.1667em\lower.7ex\hbox{E}\kern-.125emX}}

\begin{document}

%\title{\textbf{\huge{Usability of Mobile Forms for Health Needs Assessment\\
\title{\textbf{\huge{Viability of Mobile Forms for Population Health Surveys\\
in Low Resource Areas}}\\

{
\thanks{* Denotes equal contribution.}
}
}

% \author{\IEEEauthorblockN{Alexander Davis\textsuperscript{*}}
% \and
% \IEEEauthorblockN{Aidan Chen\textsuperscript{*}}
% \and
% \IEEEauthorblockN{Aidan Chen\textsuperscript{*}}
% \and
% \IEEEauthorblockN{Aidan Chen\textsuperscript{*}}
% }

\author{\IEEEauthorblockN{Alexander Davis\textsuperscript{*}}
\IEEEauthorblockA{\textit{Los Gatos High School} \\
a1a2davis@gmail.com}
\and
\IEEEauthorblockN{Aidan Chen\textsuperscript{*}}
\IEEEauthorblockA{\textit{Lynbrook Highschool} \\
achen3322@student.fuhsd.org}
\and
\IEEEauthorblockN{Milton Chen}
\IEEEauthorblockA{\textit{VSee} \\
milton@vsee.com}
\and
\IEEEauthorblockN{James Davis}
\IEEEauthorblockA{\textit{University of California, Santa Cruz} \\
davis@cs.ucsc.edu}

}
% \author{\IEEEauthorblockN{Alexander Davis\textsuperscript{*}}
% \IEEEauthorblockA{\textit{Los Gatos High School} \\
% Los Gatos, CA\\
% a1a2davis@gmail.com\\
% 0009-0001-8960-1798}
% \and
% \IEEEauthorblockN{Aidan Chen\textsuperscript{*}}
% \IEEEauthorblockA{\textit{Lynbrook Highschool} \\
% Cupertino, CA\\
% achen3322@student.fuhsd.org\\
% 0009-0002-6957-2122}
% \and
% \IEEEauthorblockN{Milton Chen}
% \IEEEauthorblockA{\textit{VSee} \\
% San Jose, CA\\
% milton@vsee.com\\
% 0009-0008-4528-1631}
% \and
% \IEEEauthorblockN{James Davis}
% \IEEEauthorblockA{\textit{Dept. Computer Science Engineering} \\
% {University of California, Santa Cruz}\\
% davis@cs.ucsc.edu\\
% 0000-0002-1413-2616}

% }

% \author{
% \IEEEauthorblockA{\textit{Short Paper \#32} 
% }
% \and

% \IEEEauthorblockA{\textit{Anonymous for Review} 
% }
% \and

% \IEEEauthorblockA{\textit{Anonymous for Review} 
% }
% \and

% \IEEEauthorblockA{\textit{Short Paper \#32} 
% }
% }

% \author{\IEEEauthorblockN{Author 1\textsuperscript{*}}
% \IEEEauthorblockA{\textit{Anonymous for Review} 
% }
% \and
% \IEEEauthorblockN{Author 2\textsuperscript{*}}
% \IEEEauthorblockA{\textit{Anonymous for Review} 
% }
% \and
% \IEEEauthorblockN{Author 3}
% \IEEEauthorblockA{\textit{Anonymous for Review} 
% }
% \and
% \IEEEauthorblockN{Author 4}
% \IEEEauthorblockA{\textit{Anonymous for Review} 
% }
% }

\maketitle

\begin{abstract}
Population health surveys are an important tool to effectively allocate limited resources in low resource communities. In such an environment, surveys are often done by local population with pen and paper. Data thus collected is difficult to tabulate and analyze. We conducted a series of interviews and experiments in the Philippines to assess if mobile forms can be a viable and more efficient survey method.  We first conducted pilot interviews and found 60\% of the local surveyors actually preferred mobile forms over paper. We then built a software that can generate mobile forms that are easy to use, capable of working offline, and able to track key metrics such as time to complete questions. Our mobile form was field tested in three locations in the Philippines with 33 surveyors collecting health survey responses from 266 subjects. The percentage of surveyors preferring mobile forms increased to 76\% after just using the form a few times.  The results demonstrate our mobile form is a viable method to conduct large scale population health surveys in a low resource environment.  
\end{abstract}


\section{Introduction}
%{1 Need for health care population surveys in Philippines. just setting the context for the work in general}
There is an unmet need for medical care in low resource communities and telehealth clinics have the potential to reach patients in remote regions with insufficient coverage~\cite{Dorsey16}. Our team has operated a series of free clinics with both in person and telehealth physicians serving low income populations in the Philippines with our partner organization Gawad Kalinga. Since Gawad Kalinga builds free housing in ten thousand locations across the Philippines, it can reach over one million households and mobilize many volunteers. Gawad Kalinga expressed interest in opening free clinics in all of its locations. 

%{2 Motivation for the work: We want 1M surveys, 10k locations. }

In order to understand the target population and how best to serve them, we will perform a large set of surveys.  Figure~\ref{clinic_context} provides an example of one community and our free clinic. The figure shows images of a street scene of the community, the interior of one of the health clinics, a remote doctor seeing a patient using telehealth, and a remote eye exam. 

%  \red{2 Alternate ways to solve the problem of taking surveys:
% . Choices are paper, tablet, phone.
% 	Tablet*10k cost too high.
% 	Paper is logistics nightmare, not feasible. Only real choice is phone input.}

A good solution to conduct a large survey in a low resource community can be challenging. Paper surveys are easy to use and record the data, but it would require a large amount of effort to transcribe information into digital format suitable for analysis~\cite{Vaish13}. Some health clinics adopt a parallel paper and digital format in an attempt to get the best of both~\cite{Gainer12}. Many organizations in affluent nations use dedicated tablets or laptops to conduct digital surveys, but in the Philippines this solution is too expensive and many people are not comfortable using such digital devices. A good platform solution to address these unique requirements is to use mobile phones already owned by the surveyors to conduct the survey.

The research aims to show that surveys conducted on mobile phones are feasible and efficient for data capture in low resource and remote regions. We conducted pilot interviews to understand user requirements, built survey forms software, ran field testing on surveys generated by the software, analyzed survey results, and performed post survey interviews.
 
%\red{3 	Pre-pilot with a sample of population and there are concerns about phone being too hard and about lack of internet. }

% \red{4
% 	To be used in our context there are Three required design attributes
% 		- Offline
% 		- Very easy
% 		- Instrumented for time measurements
% 	No existing system satisfies these constraints, so we built and offline form system which is suitable.}

Based on feedback from the pilot interviews, we determined that our mobile forms needed to satisfy three conditions. They should be extremely easy to use, allowing surveyors to complete the surveys with minimal training. They also should function properly even if the internet went out. Finally, the survey should be able to measure the time required to type and complete questions. No existing mobile forms satisfied these conditions, so we created software that can generate this type of mobile forms.
%\red{5 we perform a field study to evaluate our forms. Some short details about this field study. for example The pilot has X surveyors and Y total surveys collected. }

A field study was performed to evaluate the practicality of our mobile forms. Over three days, in three separate locations, 33 surveyors collected 20-question surveys from 266 respondents from the local population.

%\red{6 we have a post survey after the field study to gauge for user satisfaction}

After field testing was complete, we interviewed the surveyors to evaluate user experience and satisfaction. 

% \red{Last para: Contribution statement. This is the most important part of a paper. It tells the reader what you did and how it relates to all the other work out there. Its supposed to be something new that no one else has done or we do it better than they have done it. The rest of the paper tries to convince the reader that (a) it really works (b) it really is new or better. Need to fix up the words in the contributions to sound more professional}

\vspace{.1em}
The contribution of our work is:
\begin{itemize}
  \item Pilot interviews of surveyors
  \item Survey forms software: offline, easy, and timing aware
  \item Field testing the mobile forms based surveys 
  \item Analysis of field deployment
\end{itemize}

%  \item Demonstrating mobile form is a viable method in a low income environment

\newcommand{\SZ}{4.8cm}
\begin{figure*}[htbp]
\includegraphics[height=\SZ]{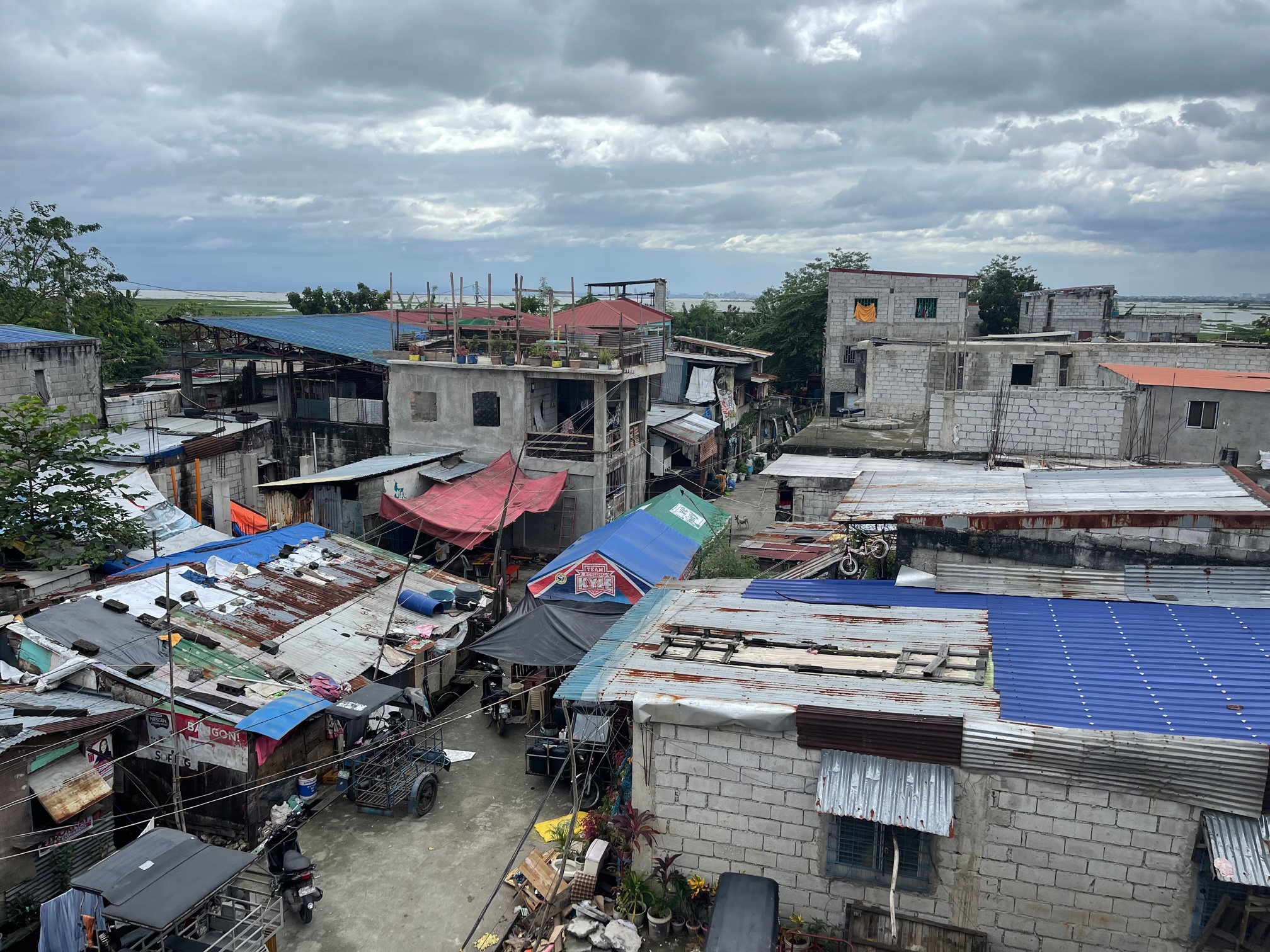}
\includegraphics[height=\SZ]{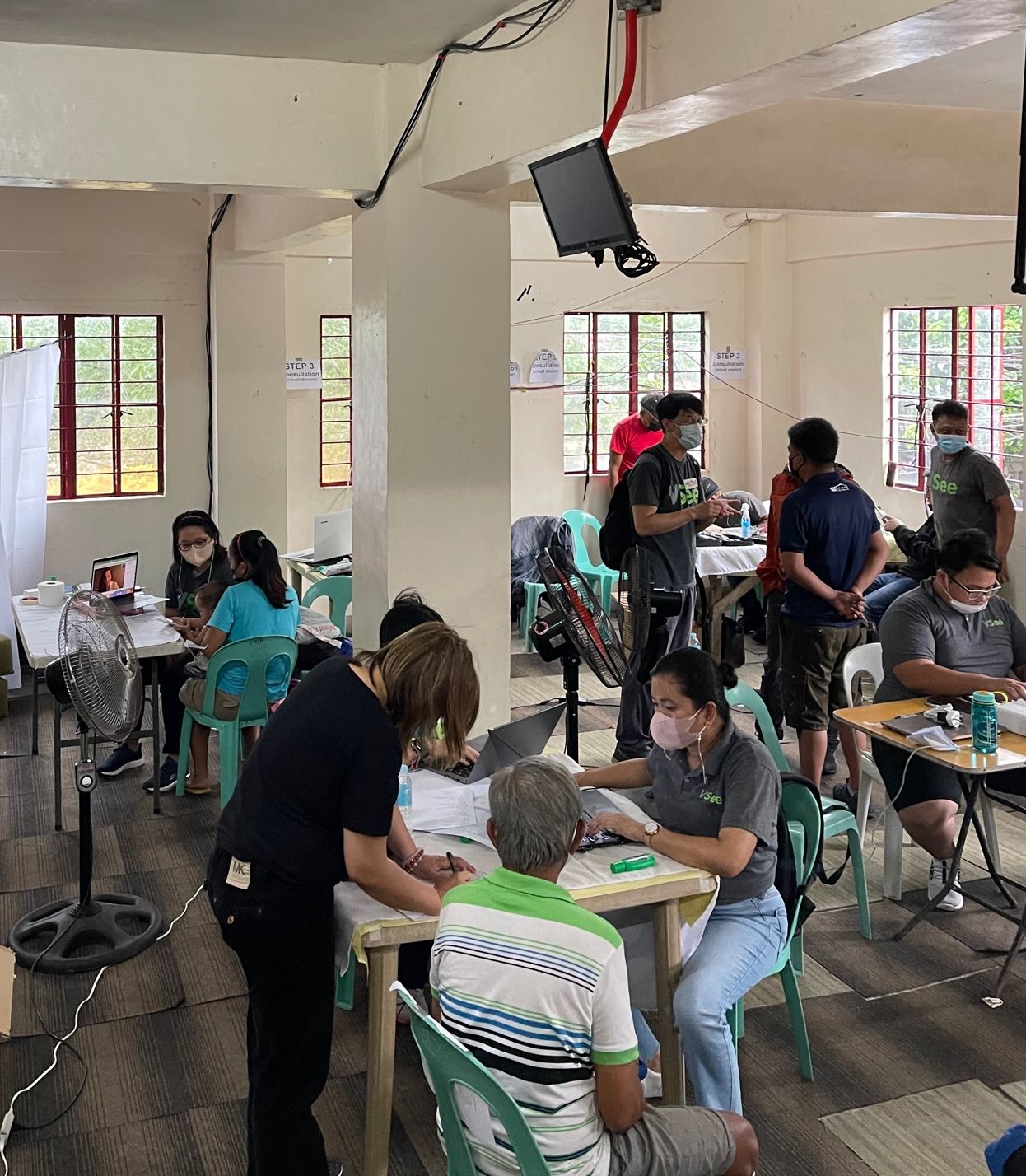}
\includegraphics[height=\SZ]{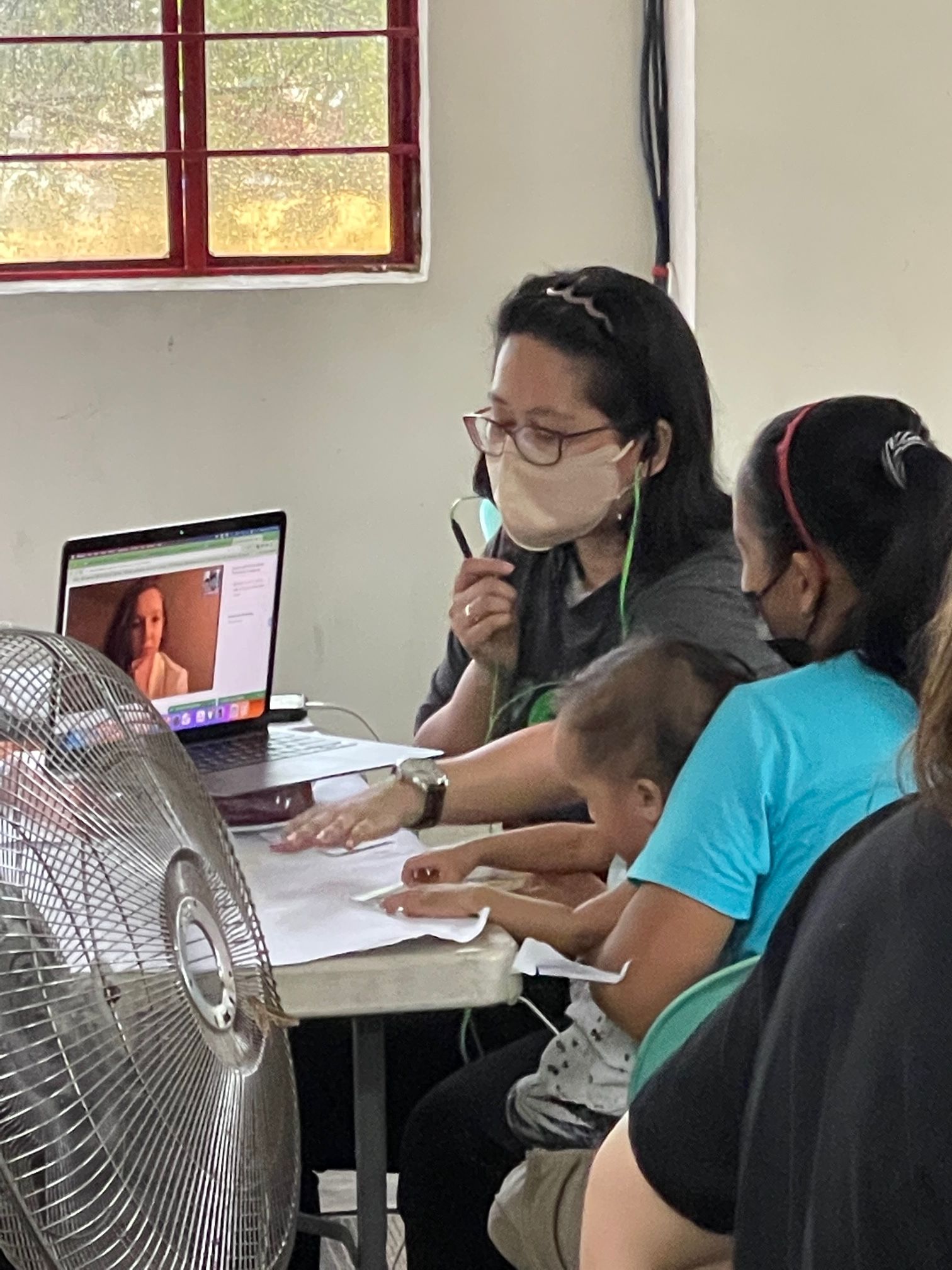}
\includegraphics[height=\SZ]{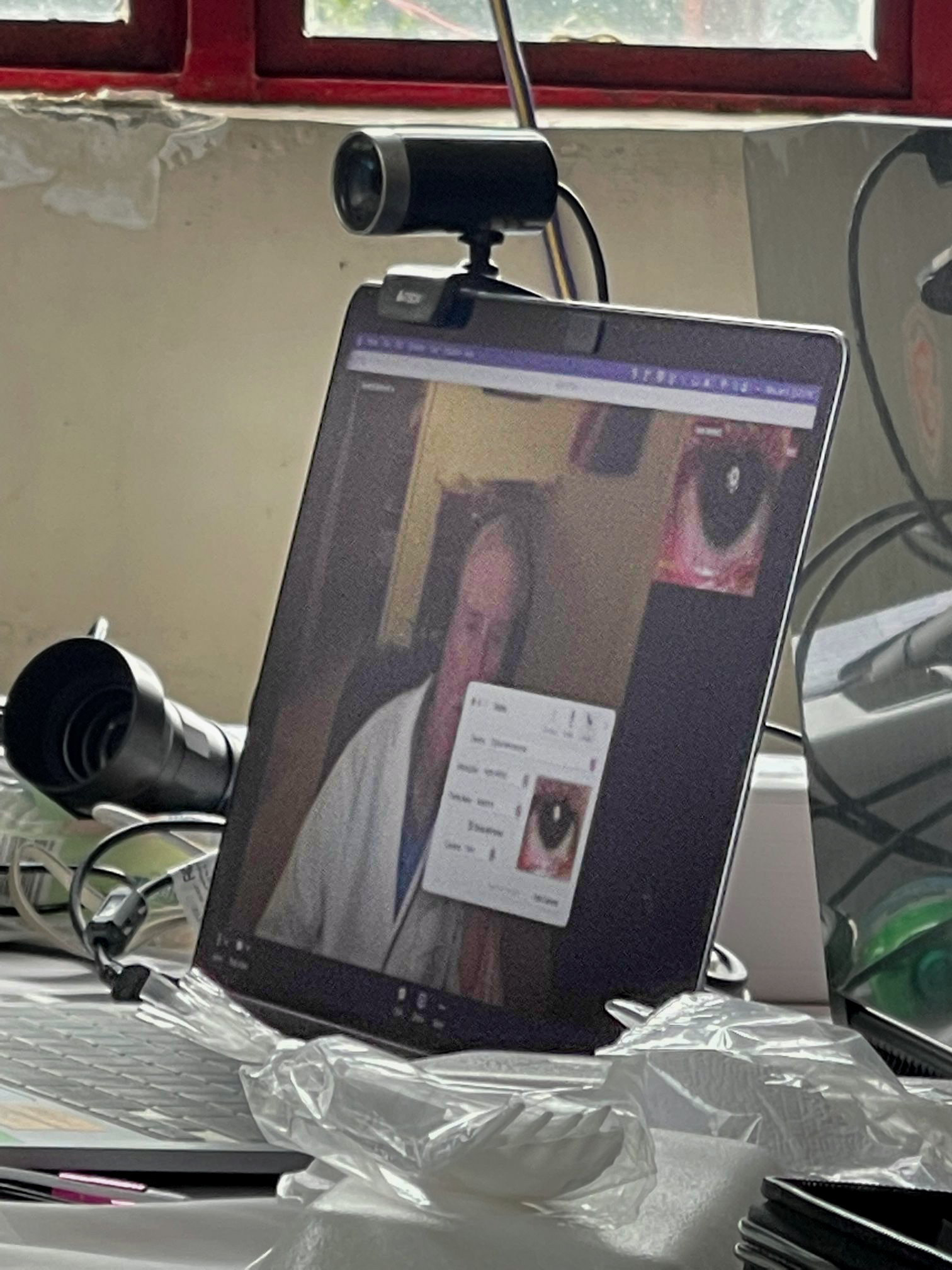}
\caption{Images from one of our telehealth clinics in Manila, Philippines. The clinics were run by VSee, a USA telehealth company. From left to right, the images portray the residential neighborhood in which the clinic took place, the interior of the clinic, a doctor (virtually on the laptop) seeing a patient, and a virtual eye examination.}
\label{clinic_context}
\end{figure*}

\section{Methodology}

%This project followed a process of pre-interview, build, test, analyze, and post-interview. We used structured pilot interviews with a set of potential surveyors to ask questions about their experience in technology and in completing health surveys. Based on the responses, customized forms software was constructed that could be used both online and offline. The forms were tested by surveyors deployed in field testing to survey members of the community about health related questions. The success rate and response time of surveyors was analyzed, and finally a post interview was performed to determine surveyor response to using the forms.

\subsection{Pilot Interviews}

We had little information on the background of our volunteer surveyors and how much knowledge they had in regards to using a phone, let alone completing a survey on one. A pilot interview was conducted to determine how comfortable the surveyors actually were with technology. {We did this by manually interviewing each surveyor and asking them questions about preference and experience.} Surveyors interviewed were chosen through convenience sampling. They were teens to seniors in their 70s from metro-Manila.

\begin{figure}[htbp]
\includegraphics[width=\linewidth]{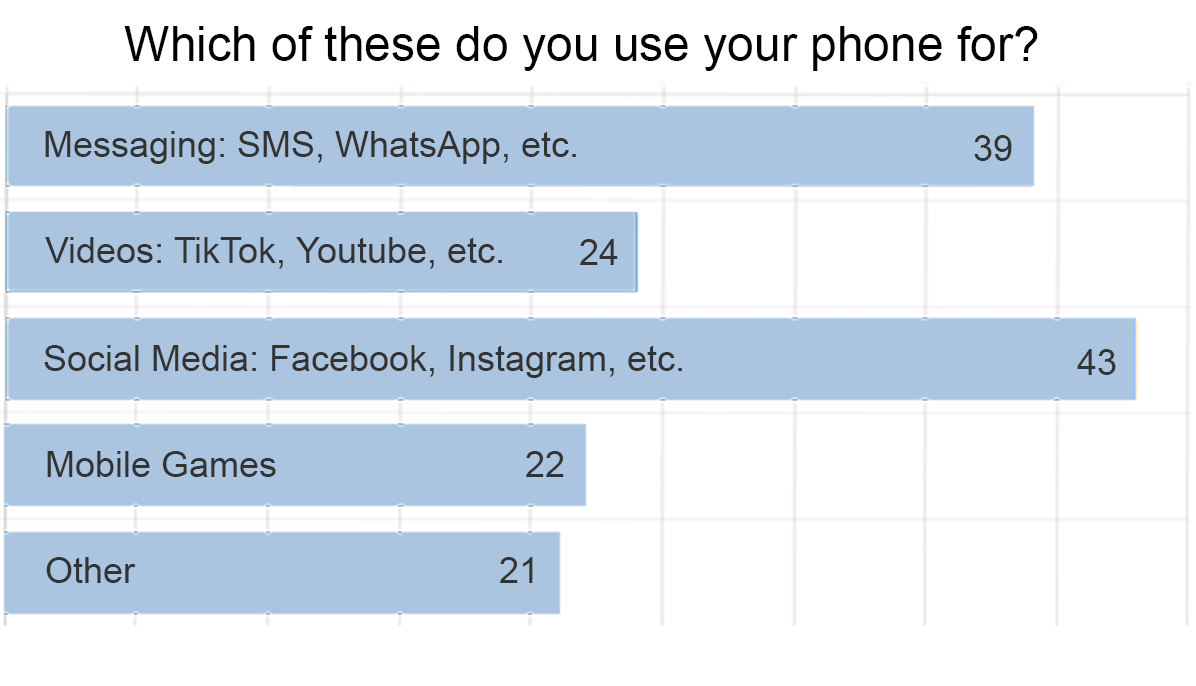}
\includegraphics[width=\linewidth]{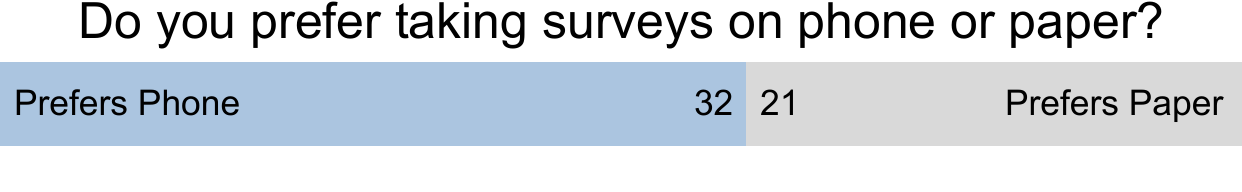}
\includegraphics[width=\linewidth]{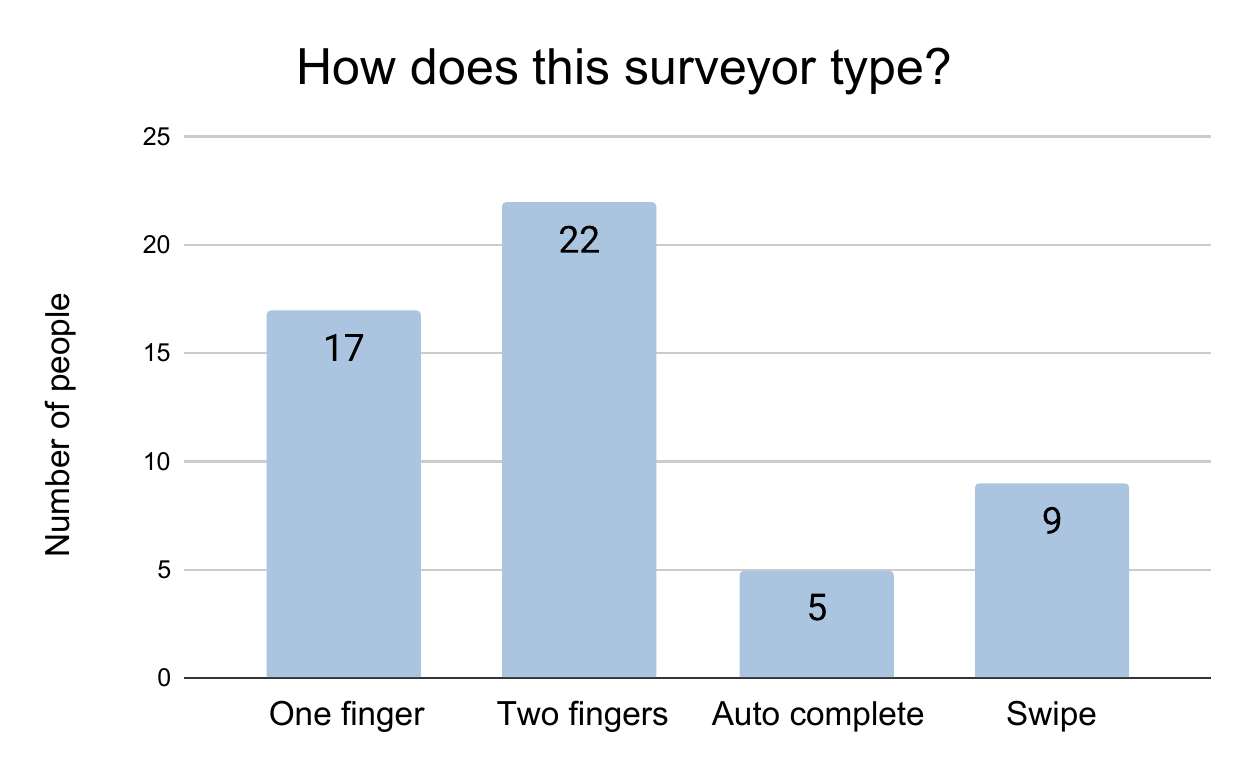}
\includegraphics[width=\linewidth, height=4em]{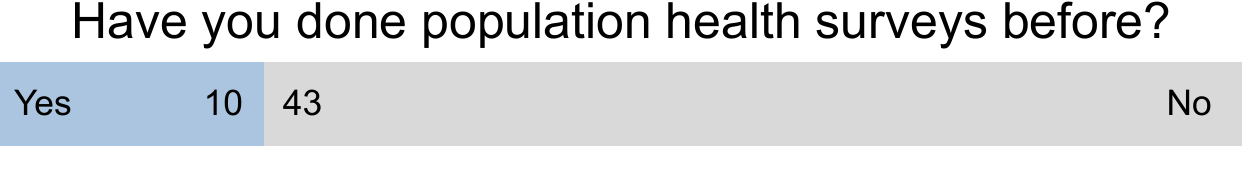}
\caption{ Pilot interviews were performed with 53 individuals to determine how comfortable volunteer surveyors were with technology. Most owned and used mobile phones for communication and entertainment purposes. However, 40\% indicated they would prefer to do surveys on paper, and we observed 32\% to type with one finger. These findings indicate that this population is not completely comfortable using technology. Only 20\% of those interviewed had any experience with population health surveys.}
\label{phonesfor}
\end{figure}

Figure \ref{phonesfor} shows a summary of results from several interview questions. Since most surveyors owned mobile phones, we first asked for what purpose their phones were used. This was a multiple-choice question that allowed surveyors to choose more than one answer. The surveyors mostly used their phones for Social Media and Messaging apps. This indicated that these surveyors were reasonably comfortable using their phones. 

We also asked if surveyors preferred completing surveys on paper or on mobile. Based on field organizers' feedback, we had anticipated the majority of surveyors preferring paper. However, only 40\% (21 out of 53) responded they had a preference for paper. When asked why they preferred paper over digital surveys, the most common answer was that paper was ``easier" or ``faster" (17 out of 21). Other answers included lack of phone ownership and concerns over poor internet access.

To understand why some are reluctant to use mobile forms, we collected data on how the surveyor type, as the style of typing may reflect familiarity and comfort with digital technology. When observing how they typed, 32\% (17~of~53) of surveyors typed with one finger, which may reflect a lack of experience and proficiency using modern digital phones. The rest typed with two fingers or used speed enhancements such as the autocomplete or swiping features on newer phones. The surveyors who typed with one finger preferred paper 53\% of the time, while those using more advanced typing methods preferred paper 30\% of the time.

Additionally, to better understand their experience with health surveys in general, we asked the surveyors if they had ever administered a population health survey before. More than 80\% responded no. 

We concluded that many of the surveyors are not proficient modern phone users and have little experience in conducting health care surveys.

\subsection{Survey Forms Software}

From our pilot interview, we learned that people had two concerns with mobile form survey. First, a mobile form would be slower and more difficult to use than paper. Second, unreliable network could render the form unusable. To evaluate mobile form's performance and improve its design, we need to track information such as how much time surveyors take to complete each question. Therefore, a suitable mobile form must be easy to learn and use, able to work offline, and capable of tracking time spent. 

While there are software solutions that can generate mobile forms for low resource environments, none satisfied all of our needs. For example EpiCollect~\cite{epicollect} lacks the ability to track the time data we desired, Google Forms does not function offline, and ODK~\cite{odk} was deemed too complex by some stakeholders. We thus developed a software that generates mobile forms that can work offline with data analysis and visualization capabilities.

%Our custom form software satisfies a set of constraints that existing systems do not: able to work offline, keeps track of timing information, and is designed to have an user-friendly interface that is easy to learn and use. The forms can also be customized by us to meet specific needs, and the software allows for easy data analysis and visualization. 

A feature of our mobile form is that it can work both online and offline. It is common to experience unstable internet in low resource environments where we collect data. At one of our previous clinic  using Google Form, we observed intermittent internet outages to be sufficiently frustrating that some surveyors started writing on paper. 

Another feature of our mobile form is that it tracks time data, so that we can study ease of use and quantify how each question will impact total effort required to collect a large number of surveys. The mobile form can measure time spent on each question, time spent typing the answer to each question, time the user spent to complete the survey, and the time that a user isn't connected to the internet. 

\begin{figure}[tbp]
 \includegraphics[trim={0 9cm 0 0 },clip,width=1\linewidth]{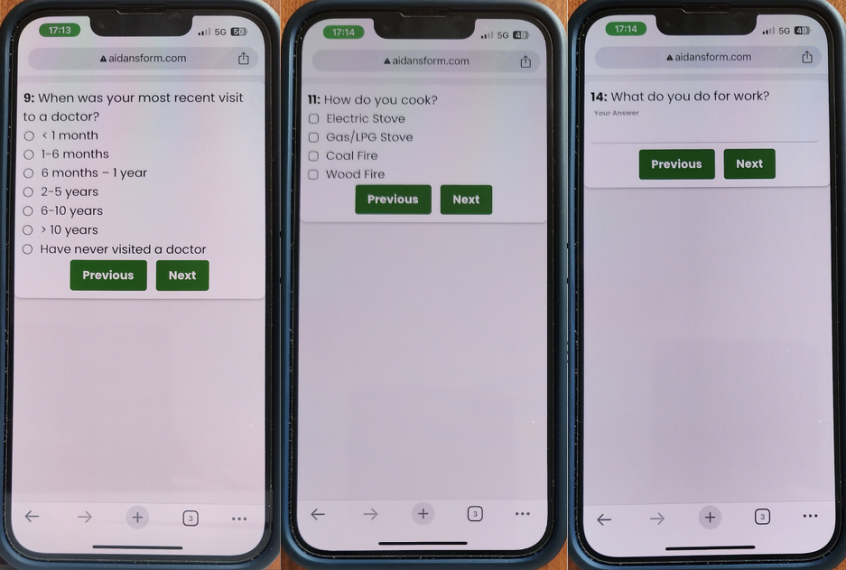}
\caption{Mobile forms were created to work offline and track the time each question took the surveyors to complete. The interface was designed to be easy to learn and use, with only one question per loaded page.}
\label{formsoftware1}
\end{figure}

To encourage surveyors to use phones, we made our mobile form as easy and intuitive as possible. We followed design recommendations for low resource populations such as minimization of visual complexity and streamlined navigation~\cite{khanna10}. Our interface is shown in Figure \ref{formsoftware1}. We presented only one question per screen to minimize scrolling, and removed all extra interface buttons that might confuse users. Our goal was to make people feel comfortable using our mobile forms and they would would prefer it over paper.

\begin{figure}[htbp]
\begin{center}
 \includegraphics[width=.60\linewidth]{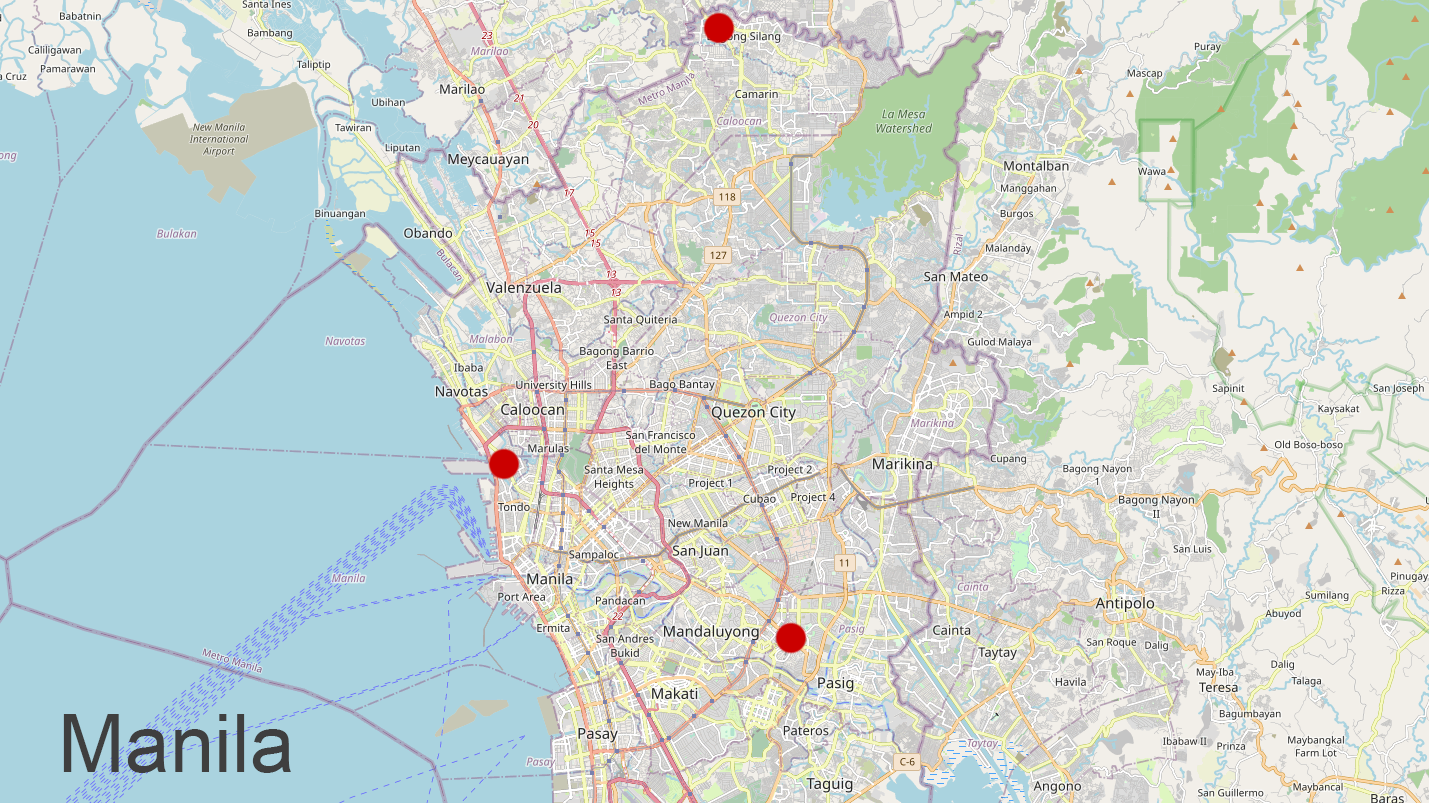}
 \end{center}
\caption{Field testing occurred at three metro-Manila sites, one in Manila City and two in the neighboring communities of Caloocan and Pasig. A total of 33 volunteer surveyors participated, collecting 266 population health surveys. }
\label{manila_map}
\end{figure}

\subsection{Field Testing}

%\red{1 We want to field test our custom software to make sure it is functional}

After designing a mobile form survey using our software, we tested its performance at three locations on three different days. The locations were in metro-Manila and ranged from an abject squatter settlement next to a trash dump to poor neighborhoods with access to water and electricity. The communities selected had between 600 and 1200 residents and were chosen to be representative of future survey sites. 
%Locations are shown in Figure~\ref{manila_map}. 

We had six of our own staff training surveyors over three days. These surveyors were volunteers from local communities, and selected to cover a wide age range. We were concerned that training might take a long time, but 96\% of surveyors completed training in 10 minutes or less.

%\red{3 How many surveyors and how many total surveys collected}

The surveyors performed a total of {266} patient surveys over three days. The survey contained 20 questions related to health and access to health care. Some example questions are``What is the closest medical center or clinic that you would go to?" and ``How do you get clean drinking water?"  The questions are a mix of textual input and multiple choice.

%The training and surveying was in the context of a field clinic in which patient health and site logistics took priority over this study. In addition, the software was being updated live and occasional database irregularities exist. Our captured data reflects these priorities, lacking entries for some actions we believe occurred. For example, while post survey interviews were collected from 33 surveyors, our database of completed survey results contains only 20 unique surveyors. In order to provide results from only surveyors we are sure participated in the study, we limit analysis to those whose names exist in both datasets.

%\red{FIGURE picture of deployment clinic and someone taking a survey on a phone.
%mention how we trained them and percentage number}

\section{Results and Discussion}

% After field testing was completed, we analyzed the data from the field tests, then gave post survey interviews so the surveyors could rate their experience with the phone survey technology. 

\subsection{Field Testing Analysis}
%\red{For each question we want to answer about typing speed, or internet available or whatever we have a paragraph that describes the question, says how we analyzed this, references a figure, and then talks about the results and conclusion of our analysis. Keep each question separated into its own paragraph. If it starts to get long we can use some bold headings to separate the different analysis we did with mini-titles}

In order to see how our surveyors actually performed when using the mobile form, we analyzed factors related to speed of completion of survey, which we hypothesized might be related to their reluctance to use mobile forms.

\begin{figure}[t]
 \includegraphics[width=\linewidth]{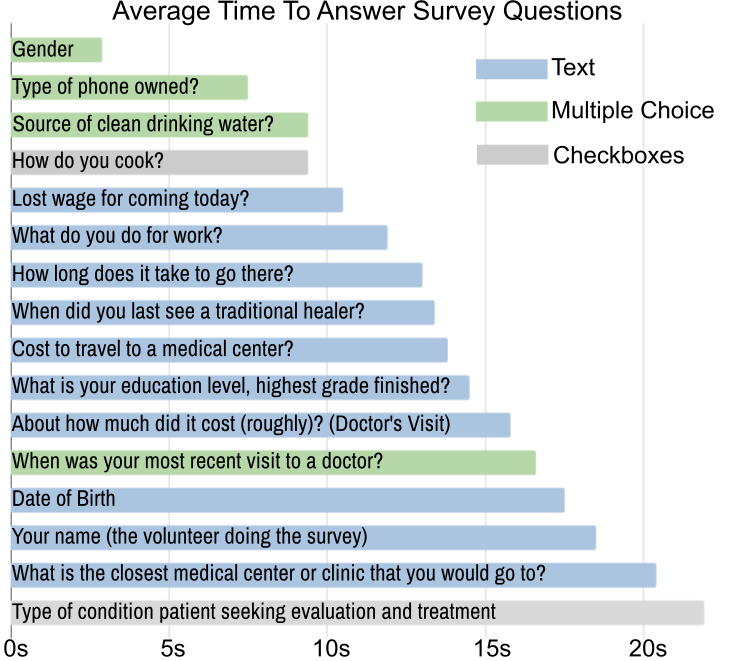}
\caption{After field testing was completed, we analyzed the time it took to complete each question. This data will be used to revise the survey to minimize the time needed to collect information. }
\label{multichoice}
\end{figure}

We wanted to find out which kind of questions were faster to complete for surveyors that had limited prior experience with mobile forms. The survey designed by our health team included both text and multiple choice questions. Figure \ref{multichoice} shows the average time taken to complete each question. Notice that on average, multiple choice questions took shorter amount of time to answer. Several of the questions could have been framed as either text or multiple choice and based on this finding we will encourage the health team to use multiple choice when possible.

\begin{figure}[t]
 \includegraphics[trim={0 0 0 0 },clip,width=\linewidth]{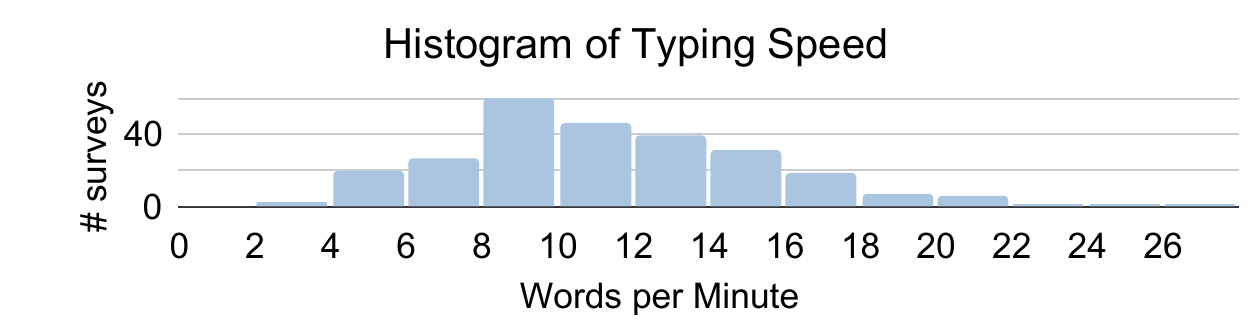}

\caption{Our mobile surveys is capable of analyzing surveyor's typing speed. The average speed was 11.5 words per minute.}
\label{typingspeed}
\end{figure}

Average typing speed in text boxes across all 266 surveys was 11.5 words per minute, however there was significant variation as shown in the histogram in Figure~\ref{typingspeed}. This is consistent with previously reported typing speeds for low resource populations~\cite{Gawade12}.

One concern raised during pilot interviews was that mobile forms would be inaccessible because there was limited internet connection. To address this concern we built mobile forms that could be used offline. To evaluate whether this issue actually occurred regularly during our field testing, a tracker was built to record the total time it took to complete the survey, and how much time on average internet was available while taking the survey. We found that 25\% of the time internet was unavailable during the survey. This indicates that a survey that can work offline is necessary.

%\red{\textbf{analysis4}... Correlation of typing speed to willingness to use paper?}

\begin{figure}[t]

\includegraphics[trim={0 0 0 0 },clip,width=\linewidth]{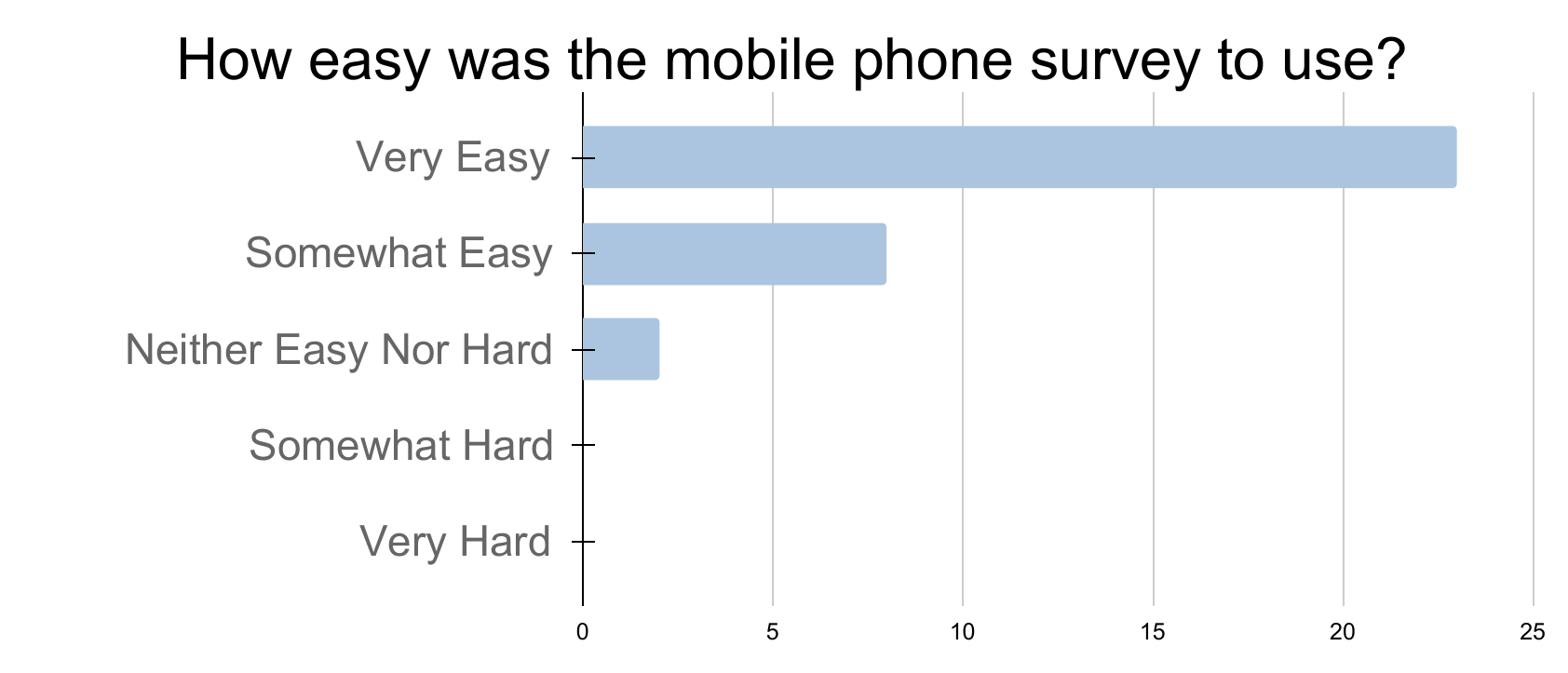}
 \includegraphics[width=\linewidth]{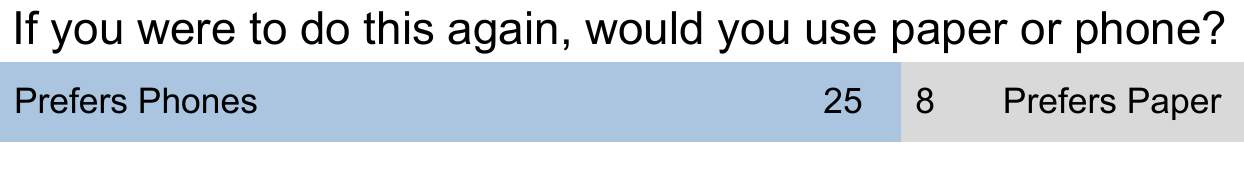}

\caption{We gave interviews after field testing to analyze surveyor experience. The majority of surveyors said that the survey was very easy to use (70\%). We also asked if they preferred mobile forms or paper after field testing, and 76\% (25 out of 33) surveyors now preferred mobile forms.}
\label{howeasy}
\end{figure}

% \begin{figure}[t]
%  \includegraphics[width=\linewidth]{PREFER WHAT V2.01.pdf}
% \caption{This is a graph that shows the proportions of what type of survey our surveyors preferred after using our custom forms survey.}
% \label{preferwhat}
% \end{figure}

%\red {1 we wanted to check if the surveyors liked our survey so we administered a post use survey. Talk about the questions, how many we did, etc}

\subsection{Post Field Testing Interviews}

We wanted to find out if the surveyors were satisfied with the mobile form they filled out. We administered post field testing interviews to all 33 surveyors.

We collected data on whether or not surveyors found the mobile form easy to use. Figure \ref{howeasy} shows 70\% of surveyors agreed the mobile form was very easy to complete. No surveyors thought it was difficult to use. We also asked surveyors to describe positive attributes about mobile forms. 30 out of 33 gave responses including the words ``faster" or ``easier", in contrast to opinions prior to using the forms. Results from these questions illustrate the effectiveness of our mobile forms.

%\red {2 results 1 from post-use}

Another question was the preference for paper or mobile forms. In the pilot interviews we found that 40\% of the surveyors preferred paper. In the post field testing interviews, we asked surveyors if they were to conduct another survey, would they use paper or mobile forms.  76\% (25 out of 33) responded that they preferred mobile over paper. Since the percentage of surveyors preferring mobile forms increased after the field testing, we hypothesize that actual experience of using mobile forms for surveying improved their opinions. Since the surveyors on average only completed the mobile form 8 times during the field testing, we believed with more usage, preference for mobile forms would continue to increase. 

%\red{ 3 results ... from post use}

% \paragraph{example figure} 

% \begin{table}[htbp]
% \caption{Table Type Styles}
% \begin{center}
% \begin{tabular}{|c|c|c|c|}
% \hline
% \textbf{Table}&\multicolumn{3}{|c|}{\textbf{Table Column Head}} \\
% \cline{2-4} 
% \textbf{Head} & \textbf{\textit{Table column subhead}}& \textbf{\textit{Subhead}}& \textbf{\textit{Subhead}} \\
% \hline
% copy& More table copy$^{\mathrm{a}}$& &  \\
% \hline
% \multicolumn{4}{l}{$^{\mathrm{a}}$Sample of a Table footnote.}
% \end{tabular}
% \label{tab1}
% \end{center}
% \end{table}

\section{Conclusion}
%\red{1 One paragraph that summarizes the contributions and findings of this paper. Shorter than the abstract but longer than the contribution sentence in the intro}
 Large scale population health surveys are essential to deploy health care resources efficiently. We investigated the feasibility of using mobile forms to conduct such surveys in low resource environment in the Philippines. Initially, Field team organizers requested to use paper to conduct the survey since many of the volunteer surveyors asked for it and there was the concern with intermittent network. However, pilot interviews revealed a mobile form was actually preferred by 60\% of the surveyors.  Based on insights gleaned from the pilot interviews, we built survey forms software that generated mobile form surveys.  We then ran field trials to test the mobile form and conducted interviews afterwards.  The health survey was successfully completed using the mobile form. The percentage of surveyors preferring mobile forms increased to 76\% after just using the form a few times.  The results demonstrate our mobile form is a viable method to conduct large scale population health surveys in this low resource environment. 

% \section*{Acknowledgment}

% \textit{Anonymous for review}.

\vspace{12pt}

\end{document}